\documentclass[aps,prb,floatfix,amsmath,amssymb,twocolumn,showpacs]{revtex4-1}
\usepackage{amsmath, amssymb, graphics, mathrsfs, bm, stackrel,bbold}
\usepackage{subfigure}
\usepackage[usenames,dvipsnames]{xcolor}
\definecolor{Highlight}{rgb}{1,1,0.75}

\allowdisplaybreaks[1]
\usepackage{graphicx}
\usepackage[bookmarks={false}, pdfauthor={Rudro Rana Biswas}, pdftitle={Semiclassical theory of viscosity in quantum Hall states}]{hyperref}
\hypersetup{colorlinks=true, linkcolor=BrickRed, citecolor=Violet, filecolor=OliveGreen, urlcolor=RoyalBlue, filebordercolor={.8 .8 1}, urlbordercolor={.8 .8 0}}
\usepackage[all]{hypcap}

\newcommand\ba{\begin{array}}
\newcommand\ea{\end{array}}
\newcommand\nn{\nonumber}

\newcommand\ri{\right}
\renewcommand\le{\left}

\newcommand{\feyn}[1]{#1\kern-0.45em/}

\newcommand\mbA{\mbs{A}}

\renewcommand\c{\psi}

\renewcommand\d{\delta}

\newcommand\D{\Delta}

\newcommand\mbe{\mbs{e}}
\newcommand\e{\epsilon}

\newcommand\mbE{\mbs{E}}

\newcommand\mbj{\mbs{j}}

\renewcommand\k{\kappa}

\renewcommand\l{\lambda}
\renewcommand\L{\Lambda}

\newcommand\p{\pi}

\newcommand\mbpp{\mbs{\p}}

\newcommand\mbr{\mbs{r}}
\newcommand\mbR{\mbs{R}}

\renewcommand\t{\tau}

\newcommand\w{\omega}

\newcommand\vx{\chi}
\newcommand\mbx{\mbs{x}}
\newcommand\X{\Xi}
\newcommand\mbX{\mbs{X}}
\newcommand\y{\eta}

\newcommand\grad{\mbs{\nabla}}
\newcommand\la{\langle}
\newcommand\ra{\rangle}
\newcommand\pd{\partial}
\newcommand\mc{\mathcal}
\newcommand\mb{\mathbb}
\newcommand\mbs{\boldsymbol}

\newcommand\mf{\mathfrak}

\newcommand\msf{\mathsf}

\begin{document}
\title{Semiclassical theory of viscosity in quantum Hall states}
\author{Rudro R. Biswas}
\email{rrbiswas@illinois.edu}
\affiliation{Department of Physics and Institute for Condensed Matter Theory, University of Illinois at Urbana-Champaign, 1110 West Green Street, Urbana, Illinois 61801-3080.}

\begin{abstract}
Quantum Hall (QH) states are predicted to display an intriguing non-dissipative stress response to a shear deformation rate, a phenomenon variously known as asymmetric or Hall viscosity, or Lorentz shear response. Just as the QH effect results from the coupling of Chern-Simons fields of the effective theory to the electromagnetic field, so also Hall viscosity is found to arise from coupling of these fields to the `metric' of the quadratic kinetic energy. In this paper I derive new physical insights for Hall viscosity by using an extended semiclassical approach to compute the conductivity of a single Landau level in a nonuniform electric field. I demonstrate that the inhomogeneity of an applied electric field is a viable experimentally tunable parameter for altering the metric, and hence creating strain in the QH state. Using these results, I argue that Hall viscosity arises from the shearing of local cyclotron orbits by the applied nonuniform electric fields.
\end{abstract}
\maketitle


Weakly interacting charged particles in magnetic fields have macroscopically degenerate (flat) energy levels, known as Landau levels (LL), a consequence of the quantization of classical cyclotron orbit motion in magnetic fields. Additionally, each fully filled LL exhibits quantization of Hall conductance, in units of $e^{2}/h$. The realization that quantized Hall conductance could be derived from quantization of the integral of the Berry curvature, over the torus of boundary phase twists applied to the quantum Hall (QH) state\cite{1982-thouless-fr,1985-niu-nr}, marked an important theoretical milestone in physics.

Recently, it has been shown that, if the quadratic kinetic energy is assumed to be coupled to a metric tensor, then the stress-energy tensor $\t$, defined with respect to variations of that metric, shows an intriguing anomalous non-dissipative longitudinal viscous response \cite{1995-avron-fk},
\begin{align}\label{eq-hallresponse}
\t_{xx} &= - \t_{yy} = \y_{H}\le(\pd u_{x}/\pd y + \pd u_{y}/\pd x\ri).
\end{align}
Here, $\mbs{u}$ is the local velocity field and $\y_{H}$ is the so-called modulus of Hall viscosity\cite{2009-read-fk}, Lorentz shear\cite{2006-tokatly-fr} or asymmetric viscosity\cite{1995-avron-fk}. Using methodology reminiscent of calculation of QH conductance, $\y_{H}$ may be calculated from the geometric response of QH wavefunctions to shear strains. Hall viscosity has received renewed interest after its calculation was extended to the fractional quantum Hall (FQH) scenario \cite{2009-read-fk, 2009-tokatly-fk, 2011-haldane-fk, 2009-haldane-rr}.

An intriguing aspect of Hall viscous response is that it also exists in non-interacting QH states. One would think that it would be straightforward to visualize the origin of this effect, in the Integer QH context, using properties of single particle cyclotron orbits. After its debut almost two decades ago, QH viscosity has been explored in a multitude of ways -- using the original method of adiabatic transport\cite{1995-avron-fk,2009-read-fk}; two-fluid hydrodynamics\cite{2006-tokatly-fr}; quantum operator methods\cite{2012-bradlyn-fk, 2010-taylor-rm}; the effective field theory of QH states coupled to a metric\cite{2012-hoyos-fk}. However, these methods are agnostic as to the integer or the fractional nature of the QH state, and therefore cannot distinguish between viscous response that is inherent to the non-commuting nature of cyclotron orbit location and velocity operators, versus the contribution from the strongly correlated nature of FQH wavefunctions.

\begin{figure}[h]
\begin{center}
\resizebox{8.5cm}{!}{\includegraphics{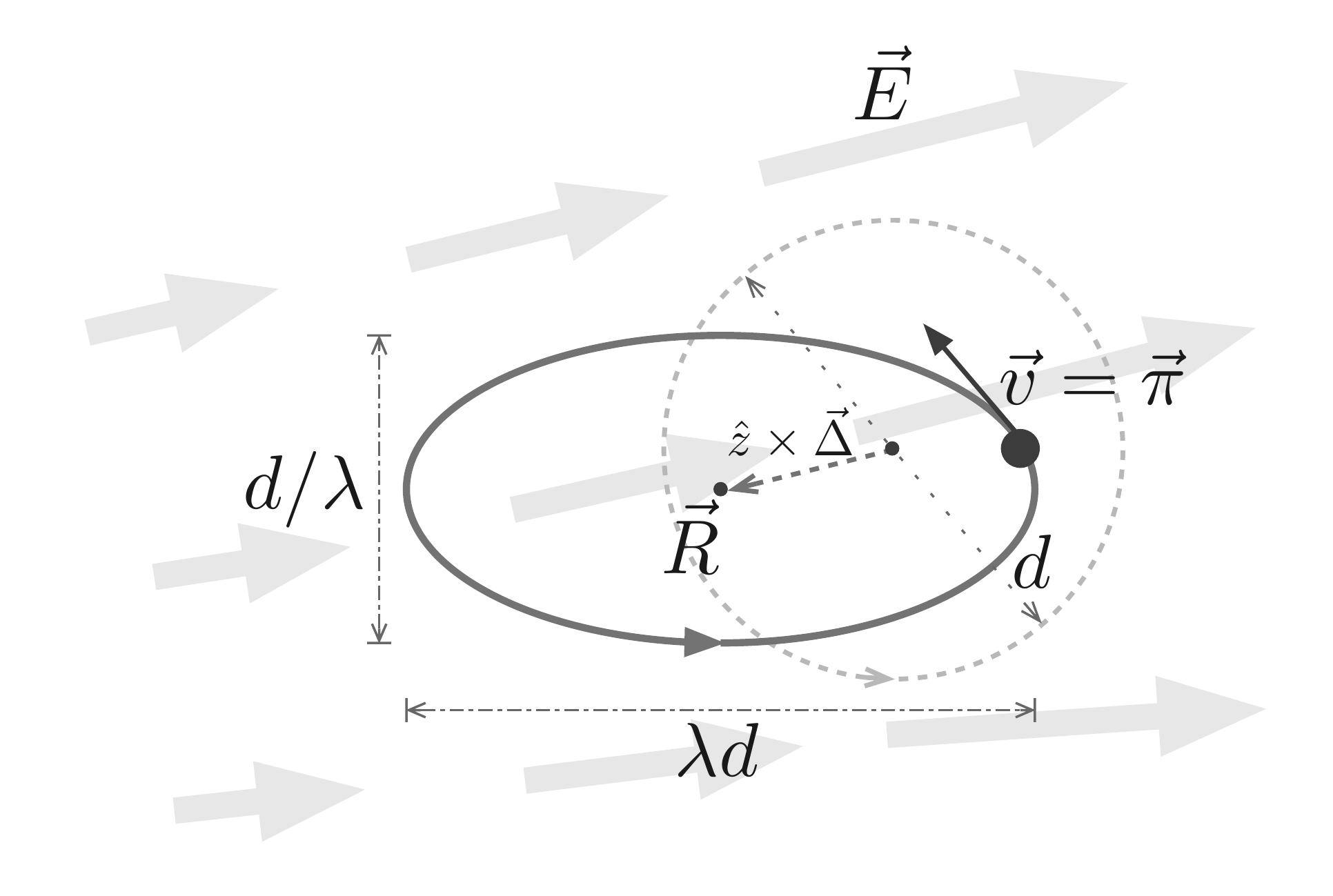}}
\caption{The cyclotron orbit (dark ellipse) in a nonuniform electric field $\mbE(\mbr)$ (gray arrows) is sheared and displaced when compared to the zero field orbit (dashed circle). The shearing parameter $\l$ and orbit orientation are determined by the eigenvalues $(\l^{2}, 1/\l^{2})$ and eigenvectors, respectively, of $\mb{g}/\sqrt{\det(\mb{g})}$, where $\mb{g}$ is the new metric (Eq.~\eqref{eq-metricchange}). The cyclotron velocity or `dynamical momentum', $\mbpp$, and `guiding center' coordinate $\mbR$ are defined in Eq.~\eqref{eq-cyclotroncoordinates}. The orbit displacement field $\mbs{\D}$ is defined in Eq.~\eqref{eq-displacementfield}.}
\label{fig-cyclotronorbit}
\end{center}
\end{figure}

\begin{figure}[t]
\begin{center}
\resizebox{8.5cm}{!}{\includegraphics{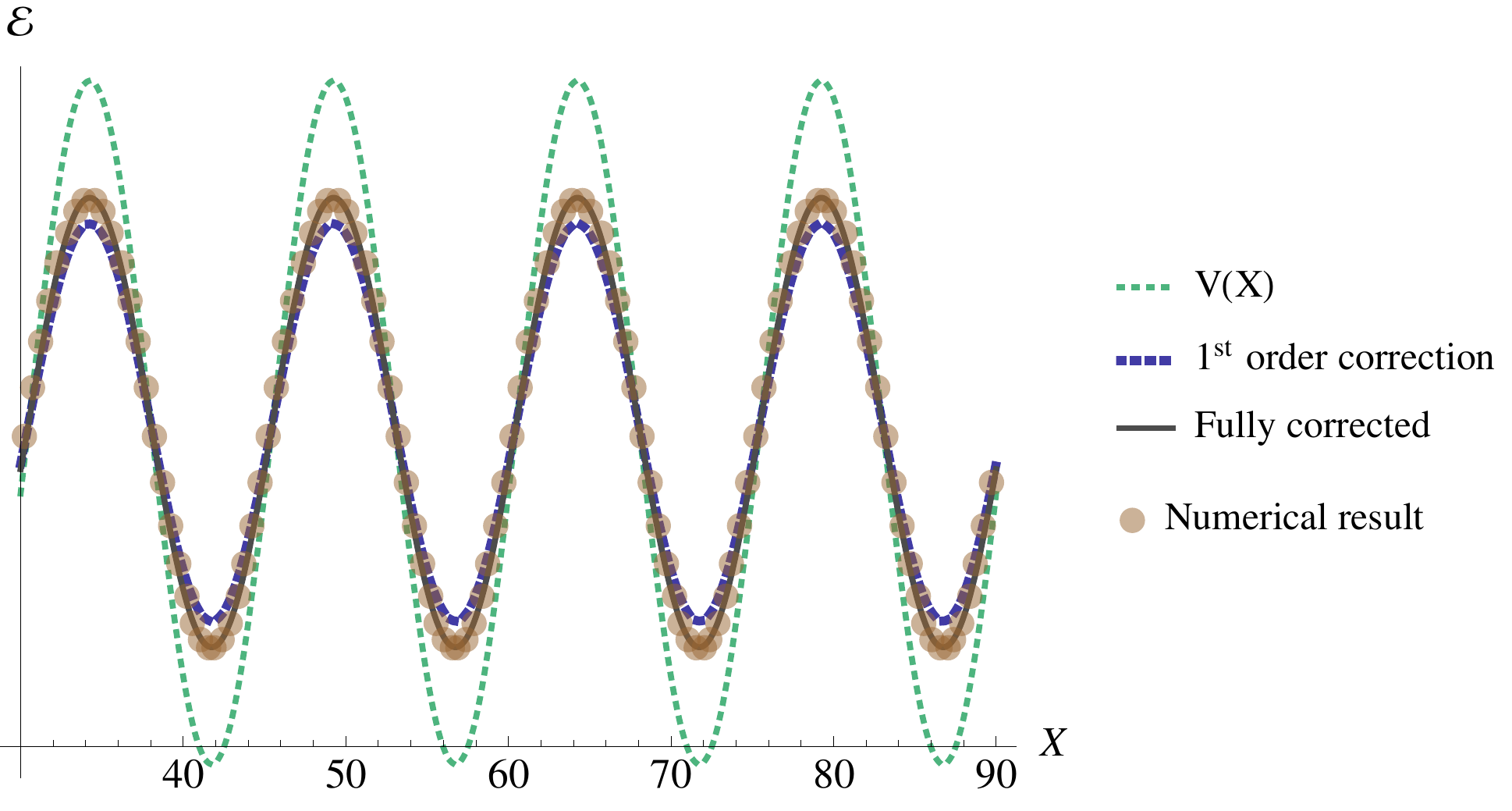}}
\caption{Variation of the lowest Landau level (LLL) energy with the location $X$ of the state, due to an applied sinusoidal background potential. In addition to the na\"ively-expected $V(X)$, the local energy is also contains corrections from the derivatives of $V$ as shown in Eq.~\eqref{eq-energysqueeze}. The `1$^{\text{st}}$ order correction' curve only uses corrections up to $V^{(2)}(X)$ which is valid for all LLs, while the full summation for $\k$, valid only for the LLL, is presented in the `Fully corrected' curve.}
\label{fig-energyvsx}
\end{center}
\end{figure}

In this paper I shall demarcate this distinction and isolate the former contribution, using the recent derivation of the effect of Hall viscosity on the conductivity of QH states in an inhomogeneous electric field\cite{1993-simon-qq}. This result was first derived by Hoyos and Son\cite{2012-hoyos-fk}(HS) and later re-derived using very different techniques\cite{2012-bradlyn-fk}. HS' first inhomogeneity correction to the uniform-field value of the current density, for a filled LL, is ($\hbar$, magnetic length and cyclotron frequency have been set to $1$)
\begin{align}\label{eq-sonresult}
\d j_{a}(\mbx) = \e_{ab}\le(-\y_{H} + \e''(B)\ri)\pd_{b}(\grad\cdot\mbE(\mbx)),
\end{align}
where $\e(B)$ is the energy density of QH states, when $\mbE=0$. I shall identify the $\e''(B)$ contribution with the $\mbE$-induced displacements of the cyclotron orbit centers, and show that the remaining Hall viscosity contribution arises because of the $\mbE$-induced shear of cyclotron orbits. To derive this, I shall revisit and extend the semiclassical theory of cyclotron motion in a LL, in the presence of a nonuniform electric field.  I shall show analytically, and verify numerically, that when the hitherto ignored mixing between LLs is correctly accounted for, the effect of a nonuniform electric field is to locally modify the cyclotron frequency\cite{1997-haldane-sf} as well as the lengthscale and guiding centers\cite{1982-halperin-oq} of the QH cyclotron orbit wavefunctions (Fig.~\ref{fig-cyclotronorbit}, Eqns.~\eqref{eq-modifiedwf}, \eqref{eq-effectivehamil2}). The local variation in cyclotron frequency (and therefore the LL kinetic energy, see Fig.~\ref{fig-energyvsx}) creates a pressure field proportional to the circulation rate of cyclotron orbit guiding centers (Eq.~\eqref{eq-pressure2}), with the constant of proportionality being, coincidentally, equal to the integer QH viscosity modulus. This pressure field modifies the well-known quantum equations of motion of the guiding center coordinates\cite{2007-jain-fk}, but is unrelated to the viscous response or any shear strain. This effect, in conjunction with the electric-field induced shearing and displacement of the orbits, is found to yield the full conductivity tensor as derived by HS.

I shall also explicitly demonstrate (Eq.~\eqref{eq-metricchange}) that the `curvature' (second derivatives) of the applied electrostatic potential modulates the metric field tensor coupled to the cyclotron kinetic energy in the LL-projected Hamiltonian, thus providing clarity to the issue of why Hall viscosity, a stress response to shear strain, is related to conductivity. This novel equivalence may be exploited to design new experimental techniques for inducing strain in QH states, thus far an impossible proposition.

\textbf{\emph{Units and notation:}} I shall use units in which the modified Planck constant $\hbar$, the electric charge magnitude $e$, the magnetic length $\ell = \sqrt{\hbar/|e B_{z}|}$ and the cyclotron frequency in the unperturbed theory $\w_{c} = |e B_{z}|/m$, $m$ being the effective electron mass, are all equal to $1$. I shall assume that $B_{z}>0$, use `blackboard bold' symbols to denote matrices\footnote{E.g., $\mb{M}$ denotes a matrix whose `$ab$' component is $M_{ab}$.}, use the notation $f^{(n)}$ to denote the $n^{th}$ derivative of $f$ and the symbol $\mf{n} = 0, 1, \ldots$ to denote the indices of the individual LLs. In order for linear response and the expansion in the wave-vector $q$ about $q=0$ to work, the applied potential $V(\mbx)$ will be assumed to be slowly varying, compared to the cyclotron gap and on the scale of the magnetic length, so that all derivatives in my reduced units are $\ll 1$. Finally, I shall ignore, across equality signs, quantities that are nonlinear in the derivatives of $V$, and (except for when explicitly stated) higher than the third order derivative of $V$.

\textbf{\emph{An easily-solved example:}} I shall first consider 2D spinless electrons in a uniform perpendicular magnetic field and a background potential $V(x)$ that is independent of $y$. This situation is simple and analytically tractable but provides all the physical ingredients necessary for the general conclusions of this paper. Translational invariance along the $y$ direction allows us to use the Landau gauge $\mbA = B_{z} x \,\hat{\mbe}_{y} \equiv x \hat{\mbe}_{y}$ and choose the energy eigenstates to be eigenstates of the $y$-momentum with momentum eigenvalue $p_{y}$. In the reduced units summarized in the previous paragraph, the Hamiltonian becomes:
\begin{subequations}
\begin{align}
\mc{H} &= \frac{p_{x}^{2} + (x - X)^{2}}{2} + V(x) \equiv \mc{H}_{0} + V(x),
\end{align}
\end{subequations}
where $X = - p_{y}$ is the `guiding center'. The eigenfunctions of $\mc{H}_{0}$ are given in terms of the simple harmonic oscillator states\footnote{Explicit wavefunctions, setting $b=1$ to match our units, may be obtained at \mbox{\url{http://dlmf.nist.gov/18.39.E5}}.} $\y_{\mf{n}}$:
\begin{align}\label{eq-unperturbedwf}
\c_{\mf{n},X} &= L_{y}^{-1/2}e^{- i X y}\y_{\mf{n}}(x-X).
\end{align}
These strip-like wavefunctions are localized around $\le\la x \ri\ra=X$ and have the positional variance
\begin{align}
\la (x - X)^{2}\ra_{\mf{n}} = \mf{n} + 1/2 \equiv 2 \vx_{\mf{n}},
\end{align}
which defines the new quantity $\vx_{\mf{n}}$. The background potential can be Taylor-expanded around the approximate location $x=X$ of the eigenstate labelled by $X$: $V(x) = V(X) + \sum_{n>0}V^{(n)}(X)(x-X)^{n}$. Using an easily-verified property of SHO wavefunctions, we can show that while considering the properties of a given LL that has been modified due to the potential $V$, to linear order in $V$, the spatial variation of the potential may be absorbed into $H_{0}$. This process is \emph{exact} for the $\mf{n}=0$ LL; for the higher LLs it works correctly to the third order derivative $V^{(3)}(X)$. The effective Hamiltonian obeyed by the states in the LL with index $\mf{n}$ is thus given by the $\mf{n}^{th}$ level of:
\begin{align}
\mc{H}_{\mf{n}} = \frac{p_{x}^{2} + (1+\k_{\mf{n}}(X))(x - X + \D_{\mf{n}}(X))^{2}}{2} + V(X)
\end{align}
with (suppressing the argument $X$ on both sides)
\begin{subequations}
\begin{align}
\D_{\mf{n}} &=\le\{\ba{ll} V^{(1)} + V^{(3)} \vx_{\mf{n}}, & \mf{n} = 0, 1\ldots\\ \sum_{m=0}^{\infty} V^{(2m+1)}\frac{\vx_{0}^{m}}{m!}, &\text{for }\mf{n}=0\text{ only} \ea\ri. \\
\k_{\mf{n}} &= \le\{\ba{ll} V^{(2)}, & \mf{n} = 0, 1\ldots \\ \sum_{m=1}^{\infty} V^{(2m)}(x_{0})\frac{\vx_{0}^{m-1}}{m!}, &\text{for }\mf{n}=0\text{ only} \ea \ri.
\end{align}
\end{subequations}
These quantities satisfy the relations (up to the relevant order, depending on $\mf{n}$, as discussed above)
\begin{align}\label{eq-relatedeltakappa}
\D_{\mf{n}} = V^{(1)} + \vx_{\mf{n}}\k_{\mf{n}}^{(1)}, \quad \D_{\mf{n}}^{(1)} = \k_{\mf{n}}.
\end{align}
The new `local' LL eigenstates labelled $(\mf{n},X)$ are thus of the form Eq.~\eqref{eq-unperturbedwf}, but with the $x$-lengthscale stretched by $\L_{\mf{n}}(X) = (1+\k_{\mf{n}}(X))^{-1/4}$ and the location shifted by $-\D_{\mf{n}}(X)$. The cyclotron frequency is also changed by a factor of $\sqrt{1+\k_{\mf{n}}(X)} = \L_{\mf{n}}(X)^{-2}$. Introducing
\begin{align}\label{eq-normal1}
\Pi= \L_{\mf{n}}(X)^{-1}(x-X+\D_{\mf{n}}(X)),
\end{align}
the new modified LL states and kinetic energies (i.e, not including $V(X)$) are
\begin{subequations}\label{eq-modifiedwf}
\begin{align}
\c_{\mf{n}, X}(x,y)&= (\L_{\mf{n}} L_{y})^{-1/2}e^{- i X y}\;\y_{\mf{n}}(\Pi),\\
\msf{E}^{K}_{\mf{n}}(X) &= \L_{\mf{n}}^{-2}\le(\mf{n} + 1/2\ri) = 2\vx_{\mf{n}}\le(1 + \k_{\mf{n}}/2\ri).\label{eq-energysqueeze}
\end{align}
\end{subequations}
I have suppressed the functional dependence of $\k_{\mf{n}}, \Pi$ and $\L_{\mf{n}}$ on $X$ in the equations above and as promised in the `Units and notation' section, ignored nonlinear corrections. These relations are at the heart of the results in this paper. The $\k_{\mf{n}}$-corrections originating from inter-LL mixing are new results and have been ignored in standard approaches for projecting to a given LL\cite{1984-girvin-pd}. In the standard approach\cite{2007-jain-fk}, the local cyclotron frequency is assumed to be unaffected and the local variation in energy is assumed to be effectively due to the potential $V(X)$ at the guiding center location\footnote{An exception is the analysis of LL levitation\cite{1997-haldane-sf}.}. I have verified this change in the cyclotron frequency via numerical calculations using a Hofstadter model\footnote{The isotropic Hofstadter model used in the figures has the inverse magnetic flux $q_{\text{Hof}}=60$ ($\ell \simeq 3.1$). The amplitude of the potential applied is $10\%$ of the LL gap.} and Figure~\ref{fig-energyvsx} displays the excellent agreement with theory, for the particular case of the lowest Landau level (LLL).

An immediate consequence of this local variation in energy is that there is a hitherto unappreciated extra force $F_{x}(X) = - \pd_{X} \msf{E}^{K}_{\mf{n}}(X) = - \vx_{\mf{n}} \k_{\mf{n}}^{(1)}(X)$ that acts on the cyclotron orbits. This modifies the drift velocity of the orbits in the $y$-direction from the uniform electric field value of $v_{y}^{(0)} = - E_{x}$ (using $E_{x}(X) = V^{(1)}(X)$)
\begin{align}\label{eq-drift1D}
v_{y}(X) = - \le(E_{x}(X) + F_{x}(X)\ri) \equiv - \D_{\mf{n}}(X).
\end{align}
In the last step, we have used Eq.~\eqref{eq-relatedeltakappa}. Using Eq.~\eqref{eq-relatedeltakappa} once more, I can re-express the excess force \emph{density} $F_{x}/(2\p) = - \pd_{X}\t_{xx}$ as arising due to a longitudinal stress $\t_{xx}$ generated from the shear rate $\pd v_{y}/\pd X$
\begin{align}\label{eq-pressure1}
\t_{xx} &= \vx_{\mf{n}}/(2\p)\;(\pd v_{y}/\pd X).
\end{align}
In this form, we may be tempted to compare this expression with that arising from Hall viscosity Eq.~\eqref{eq-hallresponse}, concluding that the Hall viscosity modulus is
\begin{align}\label{eq-hallmodulus1}
\y_{H} &= \vx_{\mf{n}}/(2\p) = (\mf{n} + 1/2)/(4\p),
\end{align}
which agrees with the known result, derived using completely different and mathematically more involved techniques\cite{1995-levay-fk}. However, this agreement seems to be a coincidence because as will be shown later in Eq.~\eqref{eq-pressure2}, $\t_{xx}=\t_{yy}$ for guiding center motion, disagreeing with Eq.~\eqref{eq-hallresponse}. The anomalous Hall response arises only when we find the \emph{local} stress, and hence sums over contributions from all cyclotron orbits passing through a given point in space. I shall explore this soon by calculating the current density response.

\begin{figure}[b]
\begin{center}
\subfigure[Hall current per state $I_{y}$ vs location $X$ of LLL states]{
\resizebox{8.5cm}{!}{\includegraphics{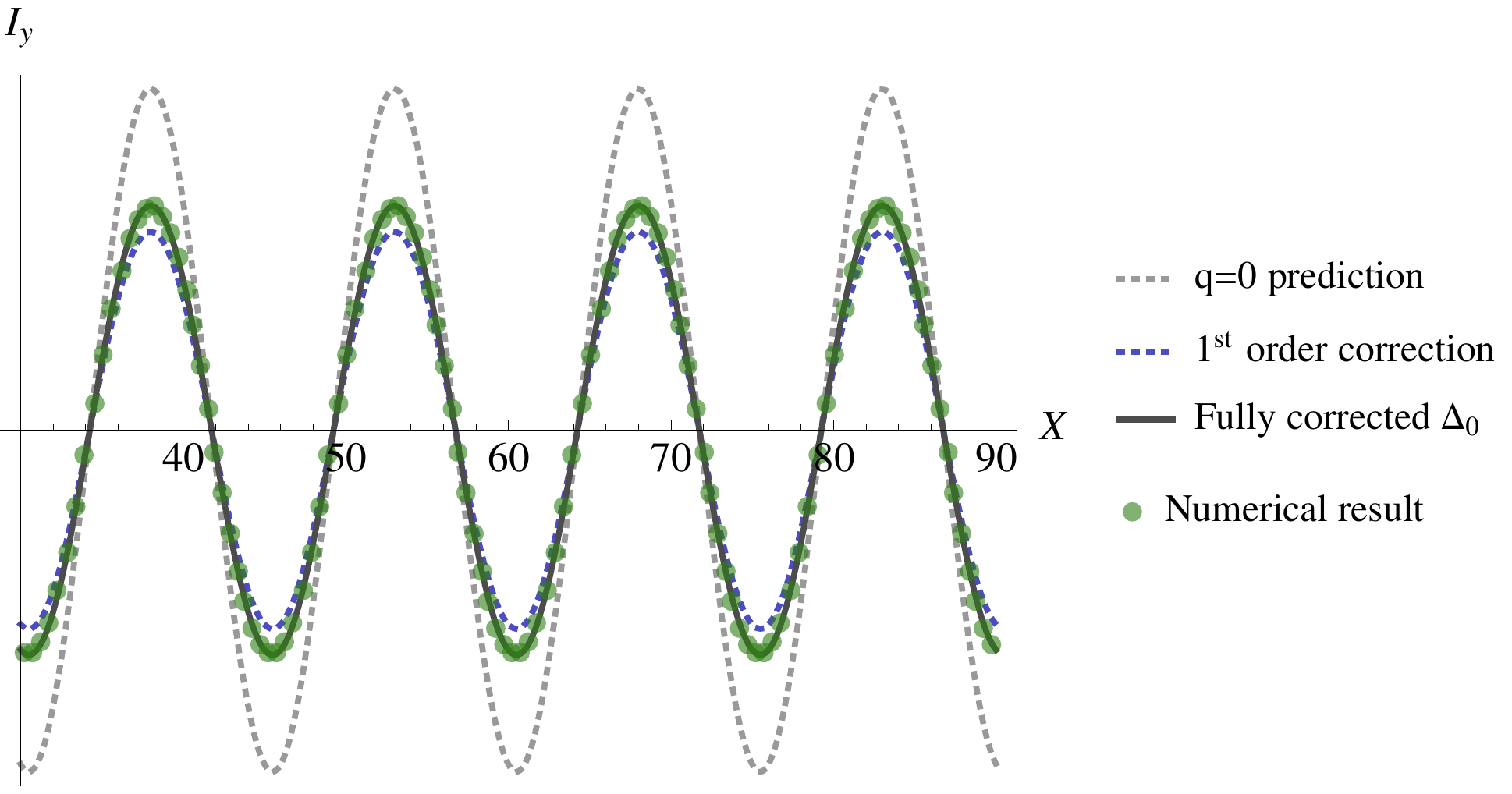}}
\label{fig-currentvsx}
}
\subfigure[Hall current density $j_{y}$ vs position $x$]{
\resizebox{8.5cm}{!}{\includegraphics{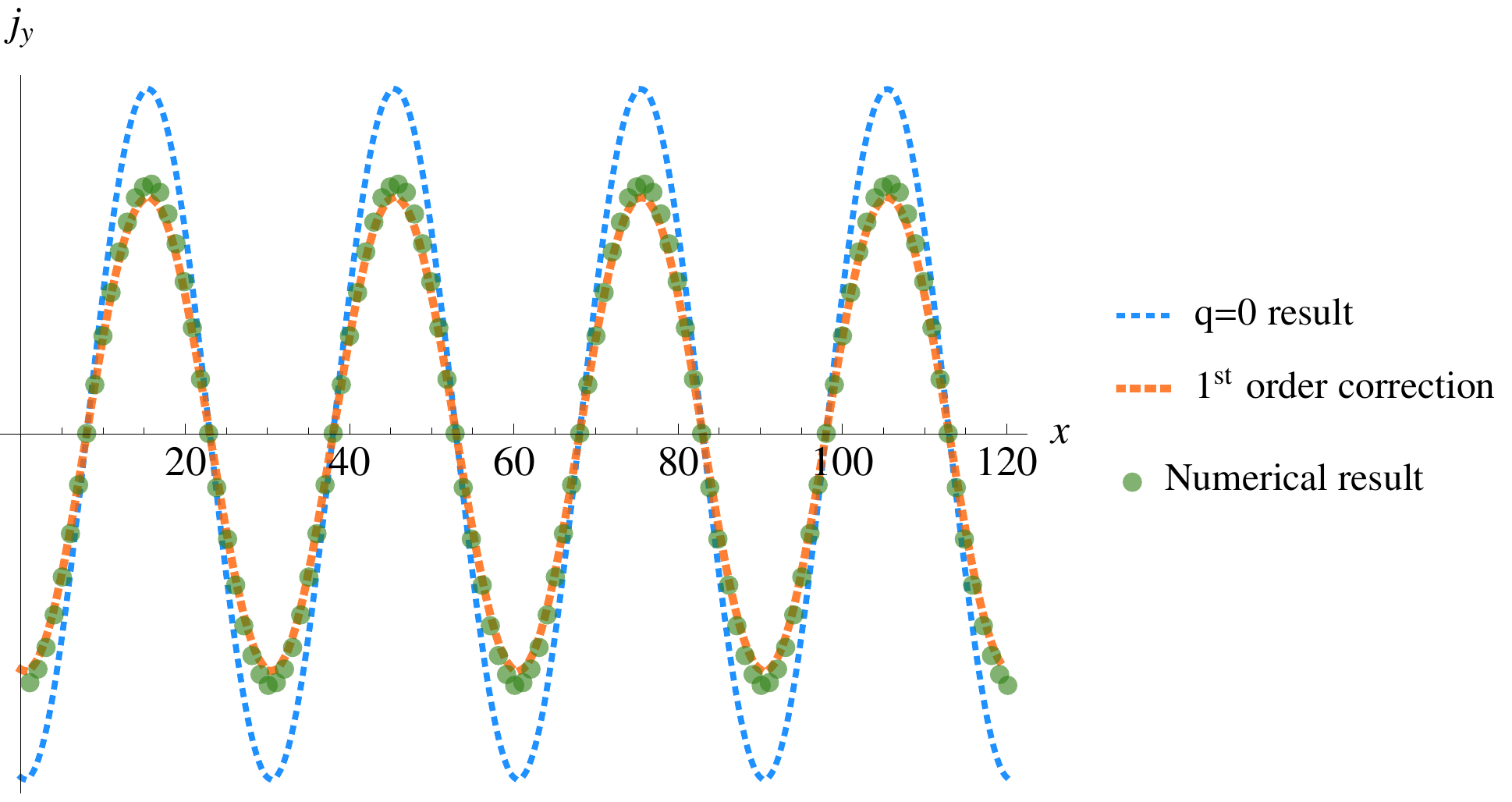}}
\label{fig-currentdensityvsx}
}
\caption{Hall current response to sinusoidally-varying potentials in the LLL -- comparison between theory (Eqns.~\eqref{eq-currentperorbit} and \eqref{eq-currentdensityresponse}) and numerical simulation using a Hofstadter model.}
\label{fig-currents}
\end{center}
\end{figure}

Using Eq.~\eqref{eq-drift1D}, I find the current $I_{y} = - v_{y}$\footnote{This is the same expression as what one may derive using the more standard definition of the current operator $\hat{I}_{y} =-e (\hat{p}_{y} + e A_{y}) = -(x - X)$, with the expectation value being consistently related to the drift velocity $ I_{y} =  - e v_{y}$.} carried by each cyclotron orbit state to be
\begin{align}\label{eq-currentperorbit}
I_{y}(X) &= \D_{\mf{n}}(X) = \le(1 + \vx_{\mf{n}} \pd_{X}^{2}\ri) E_{x}(X).
\end{align}
This prediction matches numerical calculations on the Hofstadter lattice very well, as presented in Fig.~\ref{fig-currentvsx}. This expression cannot be used to derive the spatial variation of the conductivity matrix as derived previously\cite{2012-hoyos-fk}, however, since to obtain the conductivity we need to find the local current density\footnote{I would like to thank Dam T.\ Son for first pointing this out to me.}, instead of the current carried per cyclotron orbit as derived here.

The current density operator has the form $\hat{j}_{y}(\mbx) = - (\hat{p}_{y} + A_{y}(\hat{\mbx}))\d^{2}(\hat{\mbx}-\mbx)$ and so, in the Landau gauge,
\begin{align}
\la \hat{j}_{y}(\mbx) \ra_{\mf{n},X} &= - \le(x - X\ri) |\c_{\mf{n}, X}(x,y)|^{2}.
\end{align}
Armed with this information, the current density evaluated in a full $LL$ with index $\mf{n}$ is given by
\begin{align}\label{eq-currentdensity1}
&j_{y}(\mbx) = \sum_{X} \le\la j_{y}(x) \ri\ra_{\mf{n},X}\nn\\
&= - \frac{L_{y}}{2\p}\int dX \le(\frac{x - X}{L_{y}\L_{\mf{n}}(X)}\ri) \le|\y_{\mf{n}}(\Pi)\ri|^{2} \qquad(\mbox{using Eq.~\eqref{eq-modifiedwf}})\nn\\
&\equiv (2\p)^{-1}\int d\Pi\le|\pd X/\pd \Pi\ri|\le(\D_{\mf{n}}(X) - \Pi\ri)|\y_{\mf{n}}(\Pi)|^{2}.
\end{align}
The Jacobian factor $|\pd X/\pd \Pi|$ may be ignored while considering linear response arising from the $\D_{\mf{n}}(X)$ term
\begin{align}\label{eq-currentdensitytotal1}
&j_{y}^{(A)}(x) = (2\p)^{-1}\int d\Pi\;\D_{\mf{n}}(X)|\y_{\mf{n}}(\Pi)|^{2}\nn\\
&= \frac{1}{2\p}\int d\Pi\le(\D_{\mf{n}}(x) + \D_{\mf{n}}^{(2)}(x)\frac{\Pi^{2}}{2}+\ldots\ri)|\y_{\mf{n}}(\Pi)|^{2}\nn\\
&= \frac{\D_{\mf{n}}(x) + \D_{\mf{n}}^{(2)}(x)\vx_{\mf{n}}}{2\p} = \frac{\le(1 + 2 \vx_{\mf{n}} \pd_{x}^{2}\ri) E_{x}(\mbx)}{2\p}.
\end{align}
The Jacobian, however, encodes the differential stretching and displacements of the orbits
\begin{align}
\le|\pd X/\pd \Pi\ri| = \L_{\mf{n}}(x) - \D_{\mf{n}}^{(1)}(x) - \le(2\L_{\mf{n}}^{(1)}(x) + \D_{\mf{n}}^{(2)}(x)\ri)\Pi,\nn
\end{align}
and contributes via the second term in Eq.~\eqref{eq-currentdensity1}
\begin{align}\label{eq-currentdensityB}
&j_{y}^{(B)}(x) = (2\p)^{-1}\int d\Pi \,[2\L_{\mf{n}}^{(1)}(x) + \D_{\mf{n}}^{(2)}(x)]\,\Pi^{2} |\y_{\mf{n}}(\Pi)|^{2}\nn\\
&= (-1 + 2)(\vx_{\mf{n}}/2\p) \pd_{x}^{2} E_{x}(\mbx) = (\vx_{\mf{n}}/2\p) \pd_{x}^{2} E_{x}(\mbx).
\end{align}
Combining these two contributions,
\begin{align}\label{eq-currentdensityresponse}
j_{y}(x) &= j_{y}^{(A)}(x) + j_{(2)}^{(B)}(x) = \frac{\le(1 + 3 \vx_{\mf{n}} \pd_{x}^{2}\ri)}{2\p} E_{x}(x),
\end{align}
which agrees with the result obtained by HS.

In particular, we can show that the contribution from $\L_{\mf{n}}$ and $\D_{\mf{n}}$ (and their derivatives) partition in exactly the same way as in HS' result Eq.~\eqref{eq-sonresult}; the $\L_{\mf{n}}$ contribution: the `$-1$' term in Eq.~\eqref{eq-currentdensityB}, arises from the shear of cyclotron wavefunctions (see next section) and is equivalent to what HS obtained from the Hall viscosity effect. This comparison yields the correct value of the Hall viscosity modulus $\y_{H} = \frac{\vx_{\mf{n}}}{2\p}$ (guessed earlier in Eq.~\eqref{eq-hallmodulus1}).

\textbf{\emph{Full tensorial nature of current response:}} I shall now consider the general case where the applied potential is a function of both coordinates. This discussion will closely parallel the previous simple case, when the theory is rewritten in terms of the cyclotron orbit velocity or `dynamical momentum', $\mbpp$, and the orbit `guiding center' coordinate $\mbR$, whose components and their commutation relations are:
\begin{subequations}\label{eq-cyclotroncoordinates}
\begin{align}
&\hat{\p}_{a} = \hat{p}_{a} + A_{a}(\hat{\mbx}), \hat{R}_{a} = \hat{x}_{a} - \e_{ab}\hat{\p}_{b},\\
&[\hat{\p}_{a}, \hat{\p}_{b}] = - i \e_{ab}, [\hat{R}_{a}, \hat{R}_{b}] = i \e_{ab}, [\hat{R}_{a}, \hat{\p}_{b}] = 0.
\end{align}
\end{subequations}
WLOG\footnote{The general case of an anisotropic mass tensor may be converted to the isotropic form by a linear transformation. The results we derive in this new set of coordinates need to be re-transformed back to the old coordinates.}, the electronic Hamiltonian is
\begin{align}
\mc{H} = \frac{\hat{\p}_{x}^{2} + \hat{\p}_{y}^{2}}{2} + V(\hat{\mbx}) \equiv \mc{H}_{0} + V(\hat{\mbx}).
\end{align}
The unperturbed eigenstates of $\mc{H}_{0}$ take the general form $\Psi(\p_{y}, R_{x}) = \X_{X}(R_{x}) \y_{\mf{n}}(\p_{y})$, where $\X_{X}(R_{x})$ are a set of orthonormal wavefunctions that are parametrized by some generalized index $X$. I shall use the formalism of Wigner quasiprobability distributions (WD), as that is notationally better suited for our purpose. To the lowest order in $V$, the WD $\mc{W}_{\mbX}^{(\mf{n})}$ for the guiding center coordinates factors out of the full WD and should yield a cyclotron orbit density of $(2\p)^{-1}$: $\sum_{\mbX}\mc{W}_{\mbX} = (2\p)^{-1}$. In order to find out how the cyclotron orbits are affected, we can expand $V$ about $\mbR = \mbx - \mbs{\e}\cdot\mbpp$ to obtain the following effective theory in the LL with index $\mf{n}$ (updating the notation $\k_{\mf{n}}(\mbR) = \grad^{2}V(\mbR) = \grad\cdot\mbE(\mbR)$ ):
\begin{subequations}\label{eq-effectivehamil2}
\begin{align}
\mc{H}_{\mf{n}} &= \frac{g_{ab}(\mbR)}{2}(\p_{a} + \D_{a}(\mbR))(\p_{b} + \D_{b}(\mbR)) + V(\mbR),\nn\\
g_{ab} &= \d_{ab}(1 + \k_{\mf{n}}) - \pd_{a}E_{b},\label{eq-metricchange}\\
\D_{a} &= - \e_{ab}\le[E_{b} + \vx_{\mf{n}}\pd_{b}\k_{\mf{n}}\ri].\label{eq-displacementfield}
\end{align}
\end{subequations}
This transformation shows clearly that the background potential changes the `metric' $\mb{g}$ of the high energy cyclotron motion. The latter is diagonalized by using the `normal' coordinates $\mbs{\Pi}$, satisfying $[\Pi_{a},\Pi_{b}]=-i\e_{ab}$:
\begin{align}\label{eq-normal2}
\mbs{\Pi} = \mb{\L}^{-1}\cdot(\mbpp+\mbs{\D}), \;  \mb{\L}^{\text{T}}\cdot\mb{\L} = \mb{g}/\sqrt{\det(\mb{g})}.
\end{align}
The linear transformation $\mb{\L}$ (compare Eq.~\eqref{eq-normal1}) consists of a pure rotation and a shear; the latter makes the orbits elliptical (Fig.~\ref{fig-cyclotronorbit}). The WD for the normal coordinates, $\msf{W}_{\mf{n}}(\mbs{\Pi})$, is simply that of the isotropic SHO\cite{1946-groenewold-fk} and yields the moments $\le\la \mbs{\Pi} \ri\ra = 0$, $\le\la \Pi_{a} \Pi_{b}\ri\ra = 2 \vx_{\mf{n}}\d_{ab}$.

In this effective theory, the local cyclotron frequency and hence the LL energy spacings change by a factor of $\sqrt{\det (\mb{g})} = \sqrt{1 + \k_{\mf{n}}}$. The `kinetic energy' arising from the quantization of cyclotron motion at location $\mbR$ is\footnote{The total local LL energy\cite{1997-haldane-sf} is thus $E_{\mf{n}}(\mbR) = \msf{E}^{(K)}_{\mf{n}}(\mbR) + V(\mbR)$; in usual approaches of projecting to a LL, the spatial variation of $\msf{E}^{(K)}$ is ignored.}
\begin{align}
\msf{E}^{(K)}_{\mf{n}}(\mbR) &= 2\vx_{\mf{n}}\le(1 + \k_{\mf{n}}(\mbR)/2\ri).
\end{align}
Using the expression for the drift velocity $v_{a} = \dot{R}_{a} = \e_{ab}\pd_{R_{b}}\le(\msf{E}^{(K)}_{\mf{n}} + V\ri)$, this local variation in the cyclotron energy gives rise to the excess pressure $P = \t_{xx} = \t_{yy}$:
\begin{align}\label{eq-pressure2}
P = - \vx_{\mf{n}}\k_{\mf{n}}/(2\p) = - (2\p)^{-1}\vx_{\mf{n}}\e_{ab}\le(\pd v_{a}/\pd R_{b}\ri).
\end{align}
This full tensorial version of the pressure response (compare Eq.~\eqref{eq-pressure1}), correct up to the second derivatives of $\mbE$, shows that it is unrelated to any shear strain or viscous response, and arises from the circulation of cyclotron orbits. The modulus of response is, coincidentally, the same as that of Hall viscosity: $\y_{H} = \vx_{\mf{n}}/(2\p)$.

The current density, $\hat{\mbj}(\mbx) = -\hat{\mbpp}\d^{2}(\hat{\mbx} - \mbx)$, evaluates to
\begin{align}
\la \hat{j}_{a}(\mbx)\ra &= \sum_{\mbX}\iint d^{2}Rd^{2}\Pi\,\mc{W}_{\mbX}^{(\mf{n})}(\mbR)\msf{W}_{\mf{n}}(\mbs{\Pi})\times\nn\\
&(-\p_{A}\d^{2}(\mbR - \mb{\L}_{\mf{n}}(\mbR)\cdot\mbs{\Pi} + \mbs{\D}_{\mf{n}}(\mbR) - \mbx)).
\end{align}
The sum of $\mc{W}_{\mbX}^{(\mf{n})}(\mbR)$ over $\mbX$ yields $(2\p)^{-1}$, while the $\mbR$-integral removes the $\d$-function, modulo a Jacobian
\begin{align}
\msf{J}(\mbR) = ||\mb{J}||, J_{ab} = \pd_{R_{b}}(\mbx + \mb{\L}_{\mf{n}}(\mbR)\cdot\mbs{\Pi} - \mbs{\D}_{\mf{n}}(\mbR))_{a}.\nn
\end{align}
Using Eq.~\eqref{eq-normal2}, we can finally remove $\mbpp$ and obtain
\begin{align}\label{eq-currentdensitytotal2}
j_{A}(\mbx) &= \int\! d^{2}\Pi \,\msf{W}_{\mf{n}}(\mbs{\Pi})\,\msf{J}(\tilde{\mbR})\le(\mbs{\D}_{\mf{n}}(\tilde{\mbR}) - \mb{\L}_{\mf{n}}(\tilde{\mbR})\cdot\mbs{\Pi}\ri)_{A},\nn\\
\tilde{\mbR} &= \mbx - \e\cdot (\mbs{\Pi}(\tilde{\mbR}) - \mbs{\D}(\tilde{\mbR})).
\end{align}
At this stage we can compare with Eq.~\eqref{eq-currentdensitytotal1} and in analogy to the analysis following that, break up the current density into contributions arising from the first and second terms in the bracket:
\begin{align}
j^{(A)}_{a} = -\frac{\e_{ab}}{2\p}(E_{b} + 2\vx_{\mf{n}}\pd_{b}(\grad\cdot\mbE)), j^{(B)}_{a} = -\frac{\e_{ab}}{2\p}\vx_{\mf{n}}\pd_{b}(\grad\cdot\mbE)\nn
\end{align}
These add to yield $j_{a} = -(2\p)^{-1}\e_{ab}[E_{b} + 3\vx_{\mf{n}}\pd_{b}(\grad\cdot\mbE)]$, which agrees with the derivation of HS.

We can now isolate the contributions arising from terms involving the matrix $\mb{\L}_{\mf{n}}$, which encodes the shear of cyclotron orbits, and from the orbit displacement field $\mbs{\D}_{\mf{n}}$. Comparing these contributions with those arising from the two terms in Eq.~\eqref{eq-sonresult}, I conclude that the Hall viscosity contribution is equivalent to the contribution from the term involving $\mb{\L}^{(1)}$, and so Hall viscosity must arise from the shear of the cyclotron orbits.

\textbf{\emph{Conclusion:}} I have thus derived and elucidated Hall viscosity response of quantized cyclotron orbits by deriving the semiclassical theory of cyclotron motion, in a given LL, in the presence of an inhomogeneous electric field. My derivation shows that the Hall viscosity response is caused by the shear of cyclotron orbit wavefunctions (Fig.~\ref{fig-cyclotronorbit}). Moreover, this effect does not enter the equations of motion of the guiding centers of these orbits\footnote{Incorporating a Hall viscous force into the guiding center drift yields the wrong tensor for the current density.}. Extending this approach to the FQH states by correctly incorporating correlations in the occupation of the guiding center states is an open problem whose solution will improve the quantum-semiclassical understanding of the universal properties of those states\cite{2001-susskind-lr,2002-fradkin-fk}. My approach also introduces the extra pressure term Eq.~\eqref{eq-pressure2} into the quantum equations of motion of cycloton guiding centers and this could be useful in deriving the correct magneto-roton spectrum in FQH states, since inter-particle interaction forces are modified by this effect\cite{2012-yang-qf}. Finally, this understanding of Hall viscosity using the basis of local cyclotron orbits may help rigorously extend the phenomenology to materials like graphene, which have a (non-quadratic) linear Dirac dispersion.

\begin{acknowledgments}
I would like to thank Dam T. Son for insightful discussions about this work. I thank Eduardo Fradkin, Ilya Gruzberg, Leo Kadanoff, Nick Read, Mike Stone and Paul Wiegmann for useful discussions. This research was supported by the Institute of Condensed Matter Theory (ICMT) at the University of Illinois at Urbana-Champaign.
\end{acknowledgments}

\end{document}